\documentstyle[preprint,tighten,aps]{revtex}
\newcommand{\be}{\begin{equation}}
\newcommand{\ee}{\end{equation}}
\newcommand{\ba}{\begin{eqnarray}}
\newcommand{\ea}{\end{eqnarray}}

\newcommand{\fr}[2]{\frac{#1}{#2}}
\newcommand{\non}{\nonumber}

\def\vec#1{{\mbox{\boldmath$#1$}}}
\newcommand{\p}{\mbox{$\vec{p}$}}

\newcommand{\k}{\mbox{$\vec{k}$}}

\newcommand{\r}{\mbox{$\vec{r}$}}

\newcommand{\lb}{\left (}
\newcommand{\rb}{\right )}
\newcommand{\la}{\left\langle}
\newcommand{\ra}{\right\rangle}

\begin{document}

\title{
\[ \vspace{-2cm} \]
\noindent\hfill\hbox{\rm } \vskip 1pt
\noindent\hfill\hbox{\rm Alberta Thy 12-01} \vskip 1pt
\noindent\hfill\hbox{\rm hep-ph/0108091} \vskip 20pt
Recoil correction to the magnetic moment of a
bound electron}

\author{ Alexander Yelkhovsky\thanks{
e-mail:  yelkhovsky@inp.nsk.su}}
\address{ Budker Institute of Nuclear Physics,
%and Physics Department, Novosibirsk University,\\
Novosibirsk, 630090, Russia\\
and\\
Department of Physics, University of Alberta,\\
Edmonton, AB, Canada T6C 2J1
\thanks{Temporary address}} 
\maketitle

\begin{abstract}
The effect of a finite nuclear mass on the magnetic moment of the
electron bound in the ground state of a hydrogen-like ion is
analyzed. Using the exact in $Z\alpha$ expression for the recoil
shift of the energy, I calculate the order $(Z\alpha)^4 m/M$
correction to the electron gyromagnetic ratio. The theoretical
uncertainty in the prediction for the experimentally interesting
carbon ion is decreased to the level of $1.2\times 10^{-9}$. This
provides a factor of two improvement in the precision of the
atomic mass of the electron.
\end{abstract}

\pacs{31.30.Jv, 32.10.Dk, 12.20.Ds}

Given a particle $X$ with the mass $m(X)$, its relative atomic mass
$A_r(X)$ is defined \cite{mt} by
\be
A_r(X) =  12 \frac{m(X)}{m(^{12}{\rm C})},
\ee
where $^{12}$C stands for a free neutral atom of carbon 12 in its
ground state. At present, the most precise value of the electron's
relative atomic mass \cite{fa95},
\be
A_r^{\rm c/c}(e) = 0.000\;548\;579\;911\;1\,(12), \label{acc}
\ee
is obtained by measuring the ratio of the cyclotron frequency of a
carbon 12 atomic nucleus $\omega_c(^{12}{\rm C}^{6+}) =
6|e\vec{B}|/m(^{12}{\rm C}^{6+})$ to the cyclotron frequency of
the electron $\omega_c(e) = |e\vec{B}|/m_e$ in the same magnetic
field $\vec{B}$.

It was noted by G. Werth \cite{pc} that the high precision
measurement \cite{we00} of the ratio $\omega_L/\omega_c(^{12}{\rm
C}^{5+})$, where $\omega_L$ is the Larmor precession frequency, in
the hydrogen-like carbon ion $^{12}{\rm C}^{5+}$ can offer an
alternative source of precise information about the electron mass.
The idea is to extract this mass from the ratio of the cyclotron
frequencies $\omega_c(^{12}{\rm C}^{5+})/\omega_c(e)$ by equating
the experimental value of the gyromagnetic ratio \cite{we00},
\be
g_{\rm exp}(^{12}{\rm C}^{5+}) = 2\frac{\omega_L}{\omega_c(e)} =
2\frac{\omega_L}{\omega_c(^{12}{\rm C}^{5+})}
\frac{\omega_c(^{12}{\rm C}^{5+})}{\omega_c(e)},
\ee
to the corresponding theoretical value $g_{\rm th}(^{12}{\rm
C}^{5+})$. Such an extraction can be done with better precision
than in (\ref{acc}) provided that both $\omega_L/
\omega_c(^{12}{\rm C}^{5+})$ and $g_{\rm th}(^{12}{\rm C}^{5+})$
has lower uncertainties.

Using the ratio $\omega_c(^{12}{\rm C}^{5+})/\omega_c(e) =
0.000\,228\,627\,210\,33\,(50)$ deduced from the result of Ref.
\cite{fa95}, the GSI-Mainz collaboration has obtained \cite{we00}
\be
g_{\rm exp}(^{12}{\rm C}^{5+}) =
2.001\;041\;596\;4\,(8)\,(6)\,(44). \label{gexp}
\ee
The first two numbers in parentheses show the uncertainties due to
statistical and possible systematic errors, respectively, and the
last number is due to the uncertainty in the ratio of the
cyclotron frequencies. The theoretical prediction for the above
gyromagnetic ratio was obtained in Ref. \cite{be00}:
\be
g_{\rm th}(^{12}{\rm C}^{5+}) =
2.001\;041\;590\,7\,(50)\,(12)\,(9). \label{gth}
\ee
Along with the free electron value known with very high precision
both experimentally \cite{gfexp} and theoretically \cite{gfth},
this result includes bound state effects which mainly contribute
to its uncertainties. Thus, the first theoretical uncertainty is
an estimate of the order $(\alpha/\pi)^2(Z\alpha)^2$ radiative
correction, the second one is due to numerical errors in the first
radiative correction calculated in \cite{be00} to all orders in
$Z\alpha$, and the third one is an estimate of the order
$(Z\alpha)^4 m_e/m(^{12}{\rm C}^{6+})$ recoil correction
\cite{be00,we00}.

The prediction (\ref{gth}) does not take into account the result
for the order $(\alpha/\pi)^2(Z\alpha)^2$ correction obtained in
Refs. \cite{eg97,cmy01} which shifts the central value of
(\ref{gth}) by $-11.3\times 10^{-10}$ and decreases the first
uncertainty to $(\alpha/\pi)^2(Z\alpha)^4 \approx 2\times
10^{-11}$. In the present work I demonstrate that the order
$(Z\alpha)^4 m_e/m(^{12}{\rm C}^{6+})$ recoil correction equals
\be
\delta^{(4)}_{\rm rec} g_{\rm th}(^{12}{\rm C}^{5+}) = -
\frac{m_e}{m(^{12}{\rm C}^{6+})}\frac{(Z\alpha)^4}{12}.
\label{grec4}
\ee
This correction shifts the theoretical prediction (\ref{gth}) by
$-1.4\times 10^{-11}$. Remaining uncertainty due to the recoil
effect can be estimated as $m_e/m(^{12}{\rm C}^{6+})
(Z\alpha)^5\ln(1/Z\alpha)/\pi \approx 0.7\times 10^{-11}$.
Including into (\ref{gth}) the results of Refs. \cite{eg97,cmy01}
and the result (\ref{grec4}), we can deduce from (\ref{gexp}) and
(\ref{acc}) the improved value of the electron's atomic mass:
\be
A_r^{\rm L/c}(e) = 0.000\;548\;579\;909\;33\,(22)\,(16)\,(32),
\label{aLc}
\ee
where two first numbers in parentheses are the statistical and
systematic uncertainties, respectively, imported from
(\ref{gexp}), and the third one is due to numerical errors of the
theoretical result from Ref. \cite{be00}. Note that only when
added linearly, the uncertainties in (\ref{acc}) and (\ref{aLc})
cover the difference between the corresponding central values.

The rest of this paper is devoted to the derivation of the ${\cal
O}((Z\alpha)^4 m/M)$ recoil correction (\ref{grec4}) arising due
to the finite electron-to-nucleus mass ratio $m/M$.

To the first order in $m/M$, dynamics of a hydrogen-like ion is
governed by the Hamiltonian
\be
H = \vec{\alpha} \lb \p_e - e\vec{A}(\r_e) \rb + \beta m
    - \frac{Z\alpha}{|\r_e - \r_N|}
    + \frac{\lb \p_N + Ze\vec{A}(\r_N) \rb^2}{2M}
    + \int d\vec{x} \frac{\vec{\cal E}^2 + \vec{\cal B}^2}{2}.
\label{ham}
\ee
Here $\vec{\alpha}$ and $\beta$ are the Dirac matrices acting on
the spin degrees of freedom of the electron with the mass $m$ and
the charge $e < 0$, while $\r_e$ and $\p_e$ are the electron
position and momentum operators, respectively. Corresponding
operators for the spinless nucleus with the mass $M$ and the
charge $Z|e|$ are $\r_N$ and $\p_N$. The electron interacts with
the structureless nucleus through the instantaneous Coulomb
potential. Both particles interact also with the transverse vector
potential
\be
\vec{A}(\vec{x}) = \frac{\vec{B}\times\vec{x}}{2}
                   + \vec{\cal A}(\vec{x}),
\ee
which consists of the classical and quantum parts. Besides the
interaction terms, the latter enters into (\ref{ham}) through
$\vec{\cal E}(\vec{x}) = -i\delta/\delta \vec{\cal A}(\vec{x})$
and $\vec{\cal B} (\vec{x}) = -i\vec{\nabla}_\vec{x} \times
\vec{\cal A}(\vec{x})$.

To find the order $m/M$ correction to the electron magnetic
moment, we have to split the Hamiltonian into an unperturbed part
and a perturbation. The former describes the system in the limit
of no recoil, $M \to \infty$, and hence depends on the position of
the electron with respect to the nucleus $\r = \r_e - \r_N$. Since
in this limit the nuclear momentum $\p_N$ drops out of the
Hamiltonian (\ref{ham}), the canonical momentum of the electron
$\p_e$ coincides with $\p = -i\partial/\partial\r$. The most
appropriate way to maintain the dependence on $\r$ in the
unperturbed Hamiltonian is to perform the unitary transformation
\cite{gh71}
\be
\psi \to U \psi, \quad U = \exp \lb i e \vec{B} \frac{\r_N \times
\r_e}{2}\rb. \label{ut}
\ee
This transformation simplifies the perturbation theory which
otherwise should include the virtual transitions into excited $P$
states of both the electron with respect to the nucleus and the
ion as a whole with respect to the origin. In fact, to the first
order in the magnetic field (sufficient for the analysis of the
magnetic moment),
\be
U \to 1 + i e \vec{B} \frac{\r_N \times \r_e}{2} = 1 + i e \vec{B}
\frac{\vec{R} \times \r}{2},
\ee
where the vector $\vec{R}\,$ points to the center of mass. Thus we
see that the unitary transformation (\ref{ut}) describes the
admixture of $P$ states induced by the external magnetic field, in
both the relative and the center of mass motions.

The transformed Hamiltonian naturally splits into the unperturbed
part and the perturbations:
\ba
U^+ H U &=& H_B + V_{\rm rad} + V_{\rm rec}; \\
H_B &=& \vec{\alpha} \lb \p - e\frac{\vec{B}\times\r}{2} \rb +
\beta m - \frac{Z\alpha}{r}
+ \int d\vec{x} \frac{\vec{\cal E}^2 + \vec{\cal B}^2}{2}, \\
V_{\rm rad} &=& - e\vec{\alpha}\vec{\cal A} (\r), \\
V_{\rm rec} &=& \frac{\lb \p_N + Ze\vec{\cal A}(0) \rb^2}{2M} -
\frac{\left\{ e\vec{B}\times\r, \p_N + Ze\vec{\cal A}(0)
\right\}}{4M}. \label{vrec}
\ea
The Hamiltonian $H_B$ describes the electron in the static
external field (Coulomb plus magnetic) and the free
electromagnetic field. In the limit $\vec{B} \to 0$ it turns into
the Dirac Hamiltonian in the Coulomb field, $H_0$. The
perturbation $V_{\rm rad}$ induces the electron interaction with
the quantized electromagnetic field, while the perturbation
$V_{\rm rec}$ gives rise to the recoil correction. We can put
$\r_N = 0$ in $V_{\rm rec}$ which already contains the overall
$1/M$. Looking for the linear response of the system to the
external magnetic field, we can also neglect the order $\vec{B}^2$
term in $V_{\rm rec}$.

At zero magnetic field, the recoil correction to the energy of the
Dirac electron in the Coulomb field was expressed in terms of the
solution to the Dirac-Coulomb problem in Refs.
\cite{sha85,ye94,pg95,sha98}:
\be
\left.\la \frac{\lb \p_N + Ze\vec{\cal A}(0) \rb^2}{2M}
\ra\right|_{B=0} = -\fr{1}{M}\int\fr{d\omega}{2\pi i}\la \psi_0
\right| \lb\p-\vec{D}_\omega\rb
            G_{E_0+\omega}\lb\p-\vec{D}_\omega \rb\left| \psi_0 \ra.
\label{purec}
\ee
The average value on the left hand side is calculated over the
fluctuations of the quantized electromagnetic field, with $V_{\rm
rad}$ included an appropriate number of times (see
\cite{sha85,ye94,pg95,sha98} for details). The average value on
the right hand side is calculated over the eigenstate of the Dirac
equation in the Coulomb field at $\vec{B} = 0$,
\be
\lb \vec{\alpha} \p + \beta m - \frac{Z\alpha}{r} \rb \psi_0(\r) =
E_0\psi_0(\r),
\label{dirpsi}
\ee
the operator $\vec{D}_\omega$ describes the transverse (magnetic)
exchange between the electron and the nucleus,
\be
\vec{D}_\omega(\r) = Z\alpha \int \fr{d\k}{(2\pi)^3} \frac{4\pi
\exp(i\vec{k}\vec{r})}{\k^2 - \omega^2} \lb \vec{\alpha} - \fr{\k
(\vec{\alpha}\k)}{\k^2} \rb,
\ee
and $G_E$ is the Green function for the Dirac equation in the
Coulomb field,
\be
\lb E-\vec{\alpha}\p -\beta m + \fr{Z\alpha}{r}\rb G_E (\r,\r') =
\delta(\r-\r'). \label{dirg}
\ee
The contour of integration over $\omega$ in (\ref{purec}) goes
from minus infinity to zero below the real axis, rounds zero from
above, and then proceeds to plus infinity above the real axis.

In Ref. \cite{ye94}, Eq.(\ref{purec}) is derived from the
requirement of the recoil correction's invariance with respect to
gauge transformations of the quantized electromagnetic field.
That derivation does not exploit a particular form of the
potential entering into the Dirac equation and hence can be
directly generalized to the present case of the Coulomb plus
magnetic field:
\be
\la V_{\rm rec} \ra = -\fr{1}{M}\int\fr{d\omega}{2\pi i}\la \psi_B
\right| \lb\vec{\pi}_+ - \vec{D}_\omega\rb
            G^B_{E_B + \omega}\lb \vec{\pi}_+ - \vec{D}_\omega \rb
            \left| \psi_B \ra.
\label{recb}
\ee
The Dirac equations for the wave and Green functions $\psi_B$ and
$G^B_{E_B + \omega}$ differ from (\ref{dirpsi}) and (\ref{dirg})
by the substitution
\be
\p \to \vec{\pi} = \p - \frac{e\vec{B}\times\r}{2}.
\ee
In Eq. (\ref{recb}), the operator
\be
\vec{\pi}_+ = \p + \frac{e\vec{B}\times\r}{2}
\ee
arises instead of $\vec{\pi}$ due to the second term in the
perturbation operator $V_{\rm rec}$, (\ref{vrec}). The same result
as (\ref{recb}) was obtained recently in Ref. \cite{sha01} in a
different manner.

An equivalent form of Eq. (\ref{recb}), which is more convenient
for the further analysis, stems from the Dirac equation:
\ba
\la V_{\rm rec} \ra &=& \la V_{\rm rec} \ra_{\rm low}
                       + \la V_{\rm rec} \ra_{\rm high}, \\
\la V_{\rm rec} \ra_{\rm low} &=& - \fr{1}{M}\int\fr{d\omega}{2\pi
i}\la \frac{\vec{\pi}_+^2
            - \left\{\vec{\pi}_+,\vec{D}_\omega\right\} }{\omega}
            \ra_B
= \fr{\la \vec{\pi}_+^2
            - \left\{\vec{\pi}_+,\vec{D}_0\right\}
             \ra_B}{2M}, \label{low} \\
\la V_{\rm rec} \ra_{\rm high}  &=& \fr{1}{M}\int\fr{d\omega}{2\pi
i}\la \lb \frac{1}{\omega}[\vec{\pi}_+,H_B] - \vec{D}_\omega\rb
            G^B_{E_B + \omega}\lb \frac{1}{\omega}[\vec{\pi}_+,H_B]
            + \vec{D}_\omega \rb \ra_B.\label{high}
\ea
Here and below $\la \ldots \ra_B$ denotes the average value over
$\psi_B$. In what follows I demonstrate that in perfect analogy
with the case of zero magnetic field, (\ref{low}) includes the
lower order ($(Z\alpha)^2$ and $(Z\alpha)^4$) contributions to the
energy\footnote{It concerns the $S$ states; for $L > 0$, there is
also a purely kinematic ${\cal O}(1)$ effect \cite{lamb}.}, while
the expansion of (\ref{high}) starts with $(Z\alpha)^5$.

The average value of the local operator (\ref{low}) can be easily
found in an analytic form. Taking the square of the operators in both
sides of the Dirac equation,
\be
\lb \vec{\alpha} (\vec{\pi}_+ - e \vec{B}\times\r) -
\frac{Z\alpha}{r} \rb \psi_B(\r) = \lb E_B - \beta m \rb
\psi_B(\r),
\ee
and taking also into account that
\be
\left\{\vec{\pi}_+,\vec{D}_0\right\} = \left\{\vec{\alpha}
\vec{\pi}_+,\frac{Z\alpha}{r} \right\} + Z\alpha \left\{
\frac{\vec{\alpha}\times\vec{n}}{2r^2}, \vec{l}_+\right\}, \quad
\vec{l}_+ \equiv \r\times\vec{\pi}_+,
\ee
we get
\ba
\la V_{\rm rec} \ra_{\rm low} =&& \frac{E_B^2 + m^2 - 2E_B \la
\beta m \ra_B}{2M} \non \\
&& + \frac{1}{M}\la e \vec{B}\vec{j} + Z\alpha e\vec{B}
(\vec{\alpha}\times\vec{n}) - Z\alpha
\frac{\vec{\alpha}\times\vec{n}}{2r^2} \vec{l}_+ -
\frac{(Z\alpha)^2}{2r^2} \ra_B, \label{ptl}
\ea
where $\vec{j} = \vec{l} + \vec{\Sigma}/2$, while $\vec{l} =
\r\times\p$. The sum of the last two terms in (\ref{ptl})
vanishes,
\be
\la - Z\alpha \frac{\vec{\alpha}\times\vec{n}}{2r^2} \vec{l}_+ -
\frac{(Z\alpha)^2}{2r^2} \ra_B = \frac{i}{4} \la \left[ H_B,
Z\alpha \{\vec{\pi}_+,\vec{n}\} \right] \ra_B = 0,
\ee
so that
\be
\la V_{\rm rec} \ra_{\rm low} = \frac{E_B^2 + m^2 - 2E_B m \la
\beta \ra_B + 2 e \vec{B}\vec{j} + 2 Z\alpha e\vec{B}
\la \vec{\alpha}\times\vec{n} \ra_B}{2M}.
\ee
To the first order in the magnetic field,
\ba
E_B &\to& E_0 - g_D \frac{e \vec{B}\vec{j} }{2m}, \\
\la \beta \ra_B &=& \la \frac{\partial H_B}{\partial m} \ra_B =
\frac{\partial E_B}{\partial m} \to \frac{1}{m} \lb E_0 + g_D
\frac{e \vec{B}\vec{j} }{2m} \rb, \\
Z\alpha e\vec{B} \la \vec{\alpha}\times\vec{n} \ra_B &\to& Z\alpha
e\vec{B} \la \vec{\alpha}\times\vec{n} \ra_0 = \frac{4
e\vec{B}\vec{j}}{3} \lb \frac{E_0^2}{m^2} - 1 \rb. \label{an}
\ea
Here $g_D$ is the bound electron gyromagnetic ratio without
radiative and recoil corrections. The average value in (\ref{an})
is calculated for the electron $S$ states by using virial
relations for the Dirac equation \cite{vir}. In the ground state,
$g_D = 2/3 ( 1 + 2\sqrt{ 1-(Z\alpha)^2 })$ \cite{breit} and $E_0 =
m\sqrt{ 1-(Z\alpha)^2 }$ (see, e.g., \cite{blp}) so that the
recoil correction to the gyromagnetic ratio derived from
(\ref{low}) reads
\be
\delta_{\rm low} g = \frac{2m}{3M} \left[ \sqrt{ 1-(Z\alpha)^2 } -
1 + 2(Z\alpha)^2 \right], \label{glow}
\ee
and agrees with the result obtained in Ref. \cite{sha01}. The
first term of its expansion in $Z\alpha$,
\be
\delta^{(2)} g = \frac{m}{M} (Z\alpha)^2,
\ee
reproduces the result of Refs. \cite{fa70,gh71,co71}.

In order to prove that the expansion of (\ref{high}) starts with
$(Z\alpha)^5$, let us first note that
\be
[\vec{\pi}_+,H_B] = [\p,H_0] = -i \frac{Z\alpha\vec{n}}{r^2},
\ee
and hence the only difference between (\ref{high}) and its
$\vec{B} = 0$ counterpart is due to first order differences
between $\psi_B$ and $\psi_0$ and between $G^B$ and $G$. For the
hard scale contribution to (\ref{high}), when $\omega \sim m$, the
interaction of the highly excited electron with the magnetic field
described by $G^B-G$, is an order $(Z\alpha)^5$ effect. In fact,
since at the hard scale $\vec{B}\times\r \sim |\vec{B}|/m$, the
inclusion of this interaction does not change the power counting
valid for the case of the zero magnetic field.

One could suppose that the $\vec{B} = 0$ power counting breaks
down at the atomic scale where $r \sim 1/(mZ\alpha)$. The following
arguments, however, show that this is not the case. If $Z\alpha
\ll 1$ we can treat atomic scale effects in the nonrelativistic
approach, i.e. expanding in $p/m \sim Z\alpha$. In particular,
such an expansion for the effective potential describing the
nonrelativistic electron interacting with the magnetic field,
starts with the Pauli term,
\be
V^{(0)} = - \frac{e}{2m} \lb \vec{l} + g_{\rm free}
\frac{\vec{\sigma}}{2} \rb \vec{B}, \label{pauli}
\ee
where $g_{\rm free} = 2 + \alpha/\pi + \ldots$ is the gyromagnetic
ratio for the free electron and $\vec{\sigma}/2$ is the
nonrelativistic spin operator. The perturbation (\ref{pauli}) is
diagonal in the basis of solutions to the Schr\"odinger equation
in the Coulomb field. Therefore, it has no effect on those
solutions as well as on the corresponding Green function and hence
on the atomic scale contribution to (\ref{high}).

The next term in the expansion
 of the effective potential, of order
$\p^2/m^2 \sim (Z\alpha)^2$, was found in \cite{eg97,cmy01}:
\ba
V^{(2)} &=& - \frac{g_{\rm free}-1}{8m^2} e
(\vec{\sigma}\times\vec{E}) (\vec{B}\times\r)  + \frac{\p^2}{4m^3}
e (\vec{\sigma}\vec{B}) \non \\
&& + \frac{g_{\rm free}-2}{8m^3} e (\vec{B}\p)(\vec{\sigma}\p) +
\frac{i}{8m^2} [H_{\rm S},\vec{\sigma}(\vec{\pi} \times
\vec{\pi}_+)]. \label{v2}
\ea
Here $\vec{E} = -Ze\vec{n}/r^2$ is the Coulomb electric field and
$H_{\rm S} = \p^2/(2m) - Z\alpha/r$ is the Schr\"odinger
Hamiltonian in the Coulomb field. As far as $V^{(2)} \sim
(Z\alpha)^2|\vec{B}|/m$ and the characteristic energy denominator
is $|E_0-E_n| \sim m(Z\alpha)^2$, this perturbation induces ${\cal
O}(|\vec{B}|/m^2)$ corrections to the Schr\"odinger wave function
and to the corresponding Green function. Hence, the first term of
the expansion in $Z\alpha$ has the same order $(Z\alpha)^5$ for
both $|\vec{B}|^0$ and $|\vec{B}|^1$ contributions to
(\ref{high}).

Thus we come to the conclusion that the order $(Z\alpha)^4$ recoil
correction to the gyromagnetic ratio of the bound electron can be
drawn from the expansion of (\ref{glow}) which results in
Eq. (\ref{grec4}).

The next step in the reduction of uncertainty of the electron mass
(\ref{aLc}) can be made by a more accurate evaluation of the first
radiative correction to the $g$ factor of the bound electron. As
far as $Z=6$ is yet much less than $1/\alpha \approx 137$, an
analytic calculation of the order $(\alpha/\pi)(Z\alpha)^4$
correction can also prove to be quite useful.

In conclusion, the new value of the electron's atomic mass is
extracted from the recently measured magnetic moment of the
electron bound in the hydrogen-like carbon ion. Due to the high
precision of this measurement, as well as of the theoretical
prediction for the gyromagnetic ratio of the bound electron, the
overall uncertainty of the new value is reduced by a factor of
about two as compared with the previous result extracted from the
ratio of the carbon nucleus cyclotron frequency to that of the
electron.

%%%%%%%%%%%%%%%%%%%%%%%%%%%%%%%%%%%%%%%%%%
%%%%%%%%%%%%%%%%%%%%%%%%%%%%%%%%%%%%%%%%%%

\subsection*{Acknowledgments}

I thank Andrzej Czarnecki for asking right questions and careful 
reading of the manuscript, and Thomas Beier, Victor Chernyak, 
Richard Ley, Vladimir Shabaev and G\"unther Werth for useful 
discussions. This
 research was partially supported by the 
Russian Foundation for Basic Research under grant number 
00-02-17646.

%%%%%%%%%%%%%%%%%%%%%%%%%%%%%%%%%%%%%%%%%%
%%%%%%%%%%%%%%%%%%%%%%%%%%%%%%%%%%%%%%%%%%
%%%%%%%%%%%%%%%%%%%%%%%%%%%%%%%%%%%%%%%%%%

\end{document}